Uncovering the inherited vulnerability of electric distribution networks


Bálint Hartmann[1,*], Tamás Soha[1], Michelle T. Cirunay[1,2], Tímea Erdei[1],

[1]HUN-REN Centre for Energy Research, KFKI Campus, Konkoly-Thege Miklós út 29–33., 1121

Budapest, Hungary

[2]Dr. Andrew L. Tan Data Science Institute, De La Salle University, 2401 Taft Ave, Malate, Manila, 1004

Metro Manila, Philippines

* Corresponding author. Tel.: +36 20 4825310. *E-mail address*: hartmann.balint@ek.hun-ren.hu


**Abstract**


Research on the vulnerability of electric networks with a complex network approach has produced

significant results in the last decade, especially for transmission networks. These studies have shown

that there are causal relations between certain structural properties of networks and their

vulnerabilities, leading to an inherent weakness. The purpose of present work was twofold: to test

the hypotheses already examined on evolving transmission networks and to gain a deeper

understanding on the nature of these inherent weaknesses. For this, historical models of a medium-

voltage distribution network supply area were reconstructed and analysed. Topological efficiency of

the networks was calculated against node and edge removals of different proportions. We found that

the tolerance of the evolving grid remained practically unchanged during the examined period,

implying that the increase in size is dominantly caused by the connection of geographically and

spatially constrained supply areas and not by an evolutionary process. We also show that probability

density functions of centrality metrics, typically connected to vulnerability, show only minor variation

during the early evolution of the examined distribution network, and in many cases resemble the

properties of the modern days.






## 1. Introduction

Power distribution networks deliver electricity to millions of consumers in each country. In order for all of this to happen safely and at the quality expected by the consumers, a large infrastructure consisting of electrical and passive devices has been constructed in the last more than hundred years. These distribution networks have a radial (or tree-like) structure, i.e. the path of electricity is designated by consecutive topological elements from the location of generation to the location on consumption. Such topologies are characterised by the low degree of redundancy and reserves, which makes them more vulnerable to disturbances.

This vulnerability is perceived by the consumers in the form of power outages. In such cases, the disturbance typically affects only one element of the path travelled by electricity, however, the above-mentioned topology can be 'disintegrated' even if only one element is switched off. In this paper, following similar methods as in [1] we show that characteristic patterns of vulnerability can be identified in the initial stages of the evolution of distribution networks, implying that vulnerability is an inherent property.

Most works focusing on the vulnerability of electric power networks examine only modern-day topologies, mostly due to the lack of historic datasets. Buzna et al. [2] studied the growth rate and the vulnerability of the French 400 kV network, identifying various phases of the evolution. Fang et al. [3] proposed a dynamic evolving model, finding that node degree distribution of the networks follow a power-law distribution and that the evolvement is significantly impacted by the node removal probability. Hartmann and Sugár [4] searched for small-world and scale-free properties in a 70-year dataset of the Hungarian transmission grid, reaching similar conclusions, i.e. the evolutionary process of grid development is driven by ever-changing underlying phenomena. Baardman examined which synthetic network models can replicate the properties of the evolving Dutch and Hungarian transmission grids [5]. A common feature of these papers is that they focus on transmission and sub-transmission grids, which show very different characteristics to distribution networks in relation to vulnerability.



Vulnerability analysis of electric power networks has received significant attention in recent years, with many studies using a complex network approach or a mix of those with the consideration of power engineering metrics [6][7]. The work of Wang et al. was along the first ones to use electrical centrality measures in relation to vulnerability of grids [8]. Dwivedi et al. propose a centrality index based on the maximum power flows and demonstrate its use on the IEEE 39-bus test system [9]. New types of centrality metrics with electrical interpretations were introduced by Chopade and Bikdash [10], using various IEEE test cases. Liu et al. use node electrical centrality, net-ability and a vulnerability index to describe the performance of IEEE 30 and 57 bus test cases [11]. However, many of such centrality measures were criticized by Verma et al. [12] to underestimate the vulnerability of the networks for not considering power flows. And while it is still under debate, whether using complex network approaches is a proper way to understand the vulnerability of electric networks [13], in our work we relied on similar toolsets to ease the comparison with earlier results of the field. Of the widely used performance metrics, we chose efficiency as the characterising metric of vulnerability. Since data from the early evolutionary period of distribution networks is scarcely available, the use of other metrics (i.e. ones relying on consumption and generation data) would be hijacked by incorrect assumptions. We note that network efficiency is a commonly used performance metric to examine modern-day networks [14]—[20].

Besides studying how the evolution of distribution networks affects their vulnerability, we were also examining the inherent nature of the network's tolerance against faults. To the authors' knowledge, this research question is also rarely represented in the literature. Galvan and Agarwal proposed a distributional metric in their paper, claiming that it can be used to identify whether the vulnerability is inherent in the size of the network or it is the results of its specific configuration [21]. Paniraghi and Maity discuss the inherent vulnerability in relation to small-world properties of power grids [22], however (as we also show), distribution networks do not resemble small-world during their evolution.



Present paper aims to contribute to this research gap by (i) studying the complex network parameters of a real-world evolving electric distribution network and by (ii) assessing how early stages of the evolutionary process affect the vulnerability of the modern-day network. The remainder of the paper is organized as follows. Section presents the network data and the metrics used for the analysis. In Section 3. results of the analysis are shown for vulnerability and centrality metrics, and these results are being discussed in their context. Finally, conclusions are drawn in Section 4.

## 2. Methods and data

### 2.1. Network data

The transmission and distribution of electricity in large volumes is ensured by the networks formed by transmission lines and substations. This hierarchically organized, interconnected system was built over the decades in such a way to minimize its investment and operational costs, to keep losses at minimum, while guaranteeing the highest possible level of operational safety for consumers. This evolutionary process, apart from interruptions caused by wars, has been going on continuously for almost 130 years, but its approach has changed from time to time, which left a mark on the structure of the network, and indirectly on its vulnerability as well.

The first cornerstones of electricity supply were the power plants established by the industrial facilities and the municipalities. The first local cooperations were created by supplying the electricity need of the neighbouring settlements from one of the industries, thus improving the economics of the power plants. This was followed by the construction of the medium-voltage network and some early elements of sub-transmission connections. In this work, we join the history of the network at this point, because the presented development process that spanned two decades starts with the post-war restoration and ends with the electrification of all settlements. For our analysis, we chose the South Transdanubia region for two reasons, (i) for the completeness of historical data, documented by Szabó [23], and (ii) for the reason that this utility region had no interconnections on



sub-transmission level during this period, which significantly affected the network development activities.

The battles through Southern Transdanubia between November 1944 and April 1945 hardly damaged the power plants, and among the major power line connections, only the one supplying the Baja region fell when the bridge carrying it was blown up. After the nationalization of the utility companies, the electrification of the settlements and industrial facilities started under a nationally unified management. The list of villages to be connected to the network was approved by the National Planning Office every year, while the completion of the work was the responsibility of the regionally authorized electricity supplier (DÉDÁSZ). Before the second world war, 312 of the 983 settlements in the region already had a network connection, but the coverage became complete in just over a decade and a half. As for industrial electrification, this mainly meant the connection of the power plants located in the Mecsek mountains (Pécsújhely, Komló, Máza) with the oil production in the Zala region, the coal mines in Mecsek mountains, the ironworks planned for the Danube, and a little later with the uranium mines.

The construction of a 35 kV main distribution network was the main direction of the development of electrical networks in this period. Between 1949 and 1952, the backbone of the later 35 kV system was built, connecting Pécs, Komló, Mohács, Nagykanizsa, Söjtör and Zalaegerszeg, among others. At the same time, settlements further away were typically connected with 20 kV lines, in all cases with a tree-like design, without the possibility of backup, while in very few places it was possible to switch between the supply areas.

As a result of all this, by the middle of the 1960s, the topology of the DÉDÁSZ network was formed in such a way that the parts serving the electrification of settlements were built according to the same technical and economic principles, and were connected by longer, direct power lines (Figure 1). Therefore, we can assume that although the size of the network increased by almost three times, its vulnerability did not fundamentally change during the period.



Using the network data, graph representations were created, where the edges represent transmission line connections, while the nodes correspond to transformer stations and power plants. In this work undirected and unweighted graphs were used due to the necessity that only limited information is available on the equipment put into operation 60-70 years ago. A similar difficulty was faced when reconstructing the topology, since in contrast to today's databases stored in geoinformatics systems, we could only rely on printed maps. Proper georeferencing of these required considerable works and accordingly, we often have to rely on approximations and expert assumptions. The network data can be accessed via the HUN-REN Data Repository Platform [24]. For the sake of comparison, we also analysed a GIS-based data of the region's network from modern days; due to ethical reasons this data is not available.

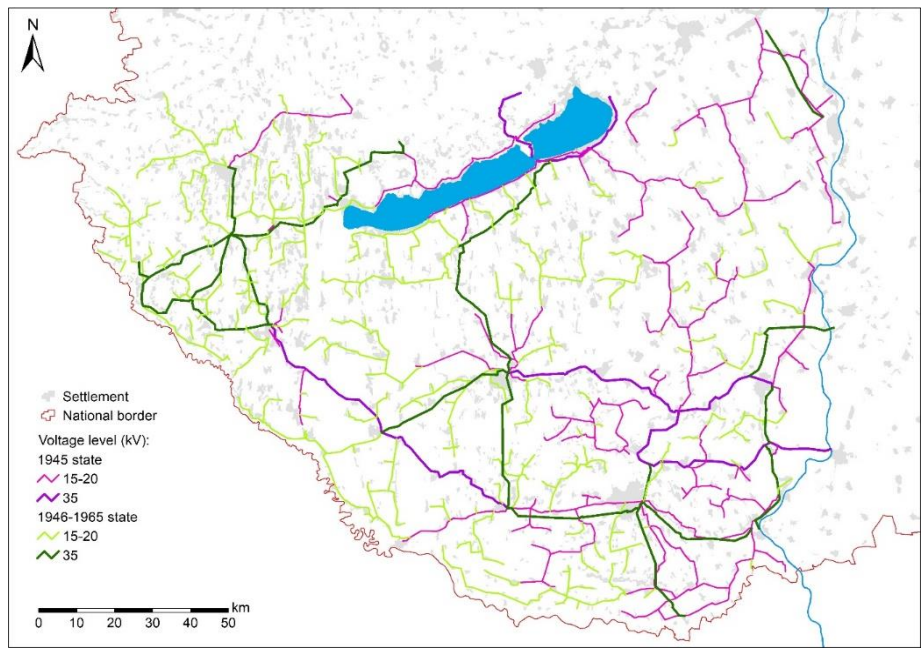

Figure 1. Development of the medium-voltage distribution network of DÉDÁSZ during the examined period

## 2.2. Metrics

The network models, described in the previous section, already provide the opportunity to examine many structural properties, which properties can usually be evaluated by calculating some complex network indicators. If the fault tolerance of the network is to be assessed, the change of these



indicators as a result of the damage provides a suitable metric. The operation of distribution networks (that is, we want to deliver electricity to all points of a supply area) is best linked to the indicator of topological efficiency, which assumes that the efficiency of power transmission between two nodes is inversely proportional to the distance between the nodes. In the following, definitions of the used metrics are presented.

To analyze the structure of the graphs, the node degree distribution, the average node degree, the link density, the average path length, the clustering coefficient, the diameter and the modularity were calculated.

The link density, $D$ is defined as the number of edges divided by the maximum possible number of edges between the number of nodes:

$$D = \frac{2E}{N(N-1)} \tag{1}$$

where $N$ is the number of nodes and $E$ is the number of edges.

The average path length, $L$ is the average number of steps along the shortest paths for all possible pairs of network vertices, defined as:

$$L = \frac{1}{N(N-1)} \sum_{j \neq i} d(i,j) \tag{2}$$

where $N$ is the number of nodes and $d(i,j)$ is the graph distance between nodes $i$ and $j$.

The clustering coefficient, $C$ is defined as:

$$C = \frac{1}{N} \sum_i \frac{2E_i}{k_i(k_i-1)} \tag{3}$$

where $N$ is the number of nodes and $E$ is the number of edges between the neighbours of i.

The modularity quotient, $Q$ is used to measure the strength of division of a network into modules and is defined as:

$$Q = \frac{1}{\langle k \rangle} \sum_{ij} \left( A_{ij} - \frac{k_i k_j}{\langle k \rangle} \right) \delta(g_i, g_j) \tag{4}$$

where $\langle k \rangle$ is the average node degree, $A_{ij}$ is the adjacency matrix and $\delta(i,j)$ is the Kronecker-delta function.



As discussed previously, efficiency was used as the characterizing performance metric. It is a measure of the network's performance, assuming that the efficiency for transmitting electricity between nodes $i$ and $j$ is proportional to the reciprocal of their distance in the graph:

$$eff = \frac{1}{N(N-1)} \sum_{j \neq i} \frac{1}{d(i,j)}$$

(5)

Vulnerability was determined as the ratio of efficiency decrease and initial efficiency ($\Delta eff./eff_0$). For each examined year (1950, 1955, 1955, 1965) 12 scenarios were defined, in which 1, 2, 5, 10, 15 and 20% of nodes or edges were removed from the network, respectively. The removed elements were selected randomly to create an even mapping of the event horizon. The number of runs was selected aiming numerical convergence of the resulting histograms, thus providing good statistical representation; each scenario was run 10 000 times. The efficiency of the network was calculated twice, once in the initial state of the network, and once after the removal of elements.

Finally, to discuss the results of the simulations, the pdfs of various centrality metrics were used. Network dynamics of the power grid are influenced by a number of factors both on the micro level (e.g. social development leading to higher per capita energy consumption), and the meso-macro level (e.g. energy and economic policies). The selected performance metric can be applied to different stages of grid evolution without it being compromised by the size of the network or the previously mentioned factors.

The degree centrality, $DC$ is a measure of local centrality that is calculated from the immediate neighborhood links of a node. It is defined as:

$$DC(v_i) = \frac{1}{N-1} \sum_{i=1}^{N} \alpha_{i,j}$$

(6)

where $N$ is the number of nodes, $\alpha_{i,j} = 1$ if there is a direct link between nodes $i$ and $j$ such that $i \neq j$.



Closeness centrality, *CC* is a measure of the average shortest distance from each node to each other node. Specifically, it is the inverse of the average shortest distance between the node and all other nodes in the network:

$$CC(i) = \frac{1}{\Sigma_i d(i,j)}$$

(7)

where $d(i,j)$ is the distance (length of the shortest path) between nodes *i* and *j*.

Betweenness centrality, BC is defined as a measure of how often a node lies on the shortest path between all pairs of nodes in a network:

$$BC(j) = \sum_{i \neq j \neq k} \frac{\sigma_{ik}(j)}{\sigma_{ik}}$$

(8)

where $\sigma_{ik}$ is the total number of shortest paths from node *i* to node *k* and $\sigma_{ik}(j)$ the number of those paths that pass through *j*.

## 3. Results and discussion

### 3.1. Network properties

The results of the structural analysis can be seen in Table 1. It can be seen that while the size of the network grew significantly between 1950 and 1965, some structural parameters remained almost unchanged. The low average node degree and the near zero clustering coefficient confirms the tree-like nature of the topology. This is also indirectly reflected through the decrease of the link density. The average path length and diameter of the network show an increase between 1950 and 1960, but then a decrease is seen in the last five years. The reason for this is an extensive 35 kV development, which connected northern and southern parts of the region. Finally, it is worth noticing that the modularity of the network stayed at a constant level with its value being very similar even in modern days. The considerable modularity means that there are dense connections between the nodes within modules but sparse connections between nodes in different modules. This confirms the hypothesis that the structure of the network was influenced by two different development



processes, the electrification of settlements (forming the modules) and the interconnection of industrial areas (connections between the modules).

Table 1. Properties of the DÉDÁSZ network

|  | N | E | D | $\langle k \rangle$ | L | C | d | Q |
|---|---|---|---|---|---|---|---|---|
| 1950 | 137 | 140 | 0.015028 | 2.0438 | 11.6844 | 0 | 28 | 0.6768 |
| 1955 | 228 | 237 | 0.009158 | 2.0789 | 15.0192 | 0.0026 | 34 | 0.6814 |
| 1960 | 311 | 321 | 0.006659 | 2.0643 | 16.4283 | 0 | 38 | 0.6756 |
| 1965 | 397 | 410 | 0.005216 | 2.0655 | 12.4654 | 0 | 28 | 0.6681 |
| Modern days | 53185 | 55176 | 0.000039 | 2.0749 | 255.4831 | 0.0021 | 747 | 0.6064 |

*3.2. Vulnerability in different periods of network evolution*

Table 2 shows the maximal values of vulnerability, given k% of nodes removed from the network at different years. It can be seen that even the removal of 1% of the nodes can significantly decrease the efficiency of the network, which is largely due its tree-like nature. Damage to the network increases as the size of the network grows, which implies that the vulnerability became worse during the years. If we compare the ratio of the maximal values for removing 1% or 20% of the nodes during the years, it is seen however that the relative tolerance to large disturbance improved. Pdfs of the results are presented in Figure 2 in detail.

Table 2. Maximal and most probable values of Δeff./eff$_0$ given k% of nodes removed

| Year | Maximal values | | |
|---|---|---|---|
|  | $k_{20\%}$ | $k_{1\%}$ | $k_{20\%}/k_{1\%}$ |
| 1950 | 0.8149 | 0.3846 | 2.1191 |
| 1955 | 0.8474 | 0.4780 | 1.7727 |
| 1960 | 0.8603 | 0.5473 | 1.5718 |
| 1965 | 0.8747 | 0.5673 | 1.5418 |



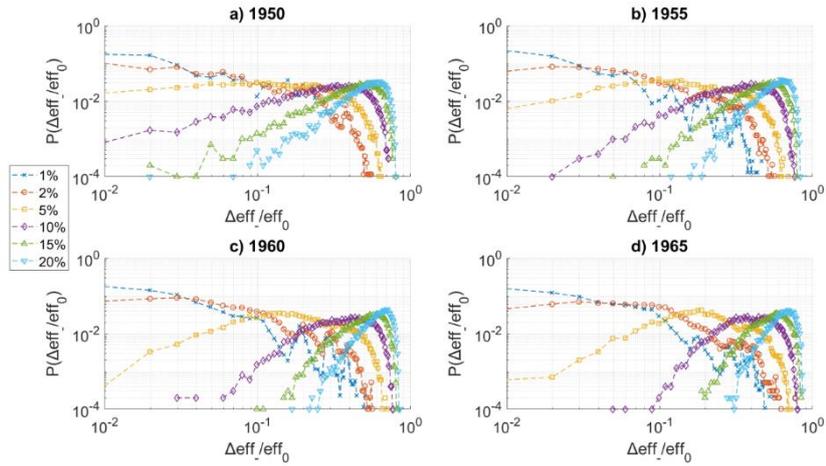

Figure 2. Decrease of network efficiency given *k%* of nodes removed in different years

Table 3. shows the maximal values of vulnerability, given k% of edges removed from the network at different years. As expected, the removal of edges causes slightly smaller damage than the removal of nodes. However, these values are still rather high, which is again related to the structure of the network; in a tree-like structure, the removal of central edges can easily decompose the network to subnetworks. Another characteristic to be highlighted is that maximal damage caused by the removal of small percentages of edges decreases from 1960 to 1965. This is the result of certain network development activities, which increased the clustering of the topology. Pdfs of the results are presented in Figure 3. in detail.

Table 3. Maximal and most probable values of $\Delta$eff./eff$_0$ given k% of edges removed

| Year | Maximal values | | |
|------|--------|--------|----------------------|
|      | $k_{20\%}$ | $k_{1\%}$ | $k_{20\%}/k_{1\%}$ |
| 1950 | 0.7892 | 0.2318 | 3.4047 |
| 1955 | 0.8144 | 0.3883 | 2.0975 |
| 1960 | 0.8435 | 0.4158 | 2.0288 |
| 1965 | 0.8624 | 0.3821 | 2.2571 |



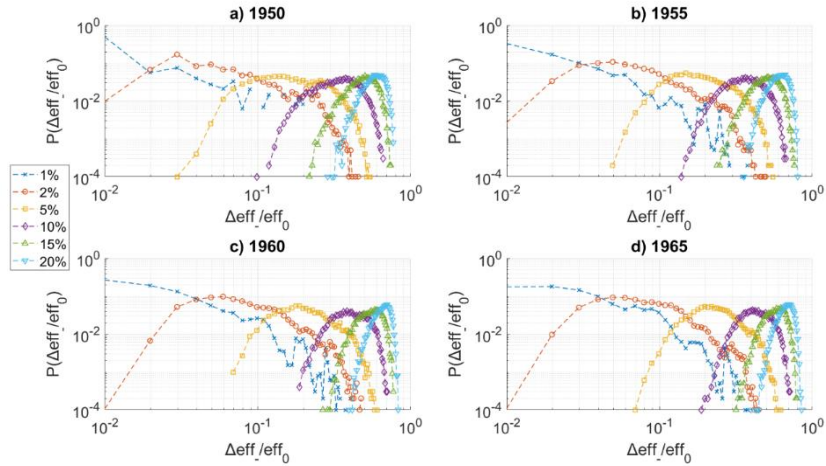

Figure 3. Decrease of network efficiency given *k%* of edges removed in different years

The results for vulnerability in different periods show that the network's resilience against removal of nodes or edges practically remained unchanged during the expansion period between 1950–1965.

*3.3. Vulnerability in case of different removals*

In the following, the vulnerability of the developing distribution network is examined from the perspective of applying the same removals to network states representing different periods. Table 4. shows the maximal values of grid vulnerability, given 1, 2, 5, 10, 15 or 20% of the nodes removed. We can identify two tendencies. First, the removal of low percentage of nodes causes significantly bigger damage in larger grids. Second, the removal of high percentage of nodes practically disables the networks ability to distribute electricity, regardless of the period under examination. These tendencies are also reflected in the ratio of damages between 1950–1965; the larger the network became, the less tolerant it became to small disturbances. Pdfs of the results are presented in Figure 4. in detail.

Table 4.  Maximal values of $\Delta$eff./$eff_0$ given k% of nodes removed

| Year | 1% | 2% | 5% | 10% | 15% | 20% |
|------|------|------|------|------|------|------|
| 1950 | 0.3846 | 0.5619 | 0.6504 | 0.7397 | 0.8138 | 0.8149 |
| 1965 | 0.5673 | 0.6364 | 0.7798 | 0.8196 | 0.8589 | 0.8747 |
| 1965/1950 | 1.4752 | 1.1326 | 1.1990 | 1.1080 | 1.0553 | 1.0733 |



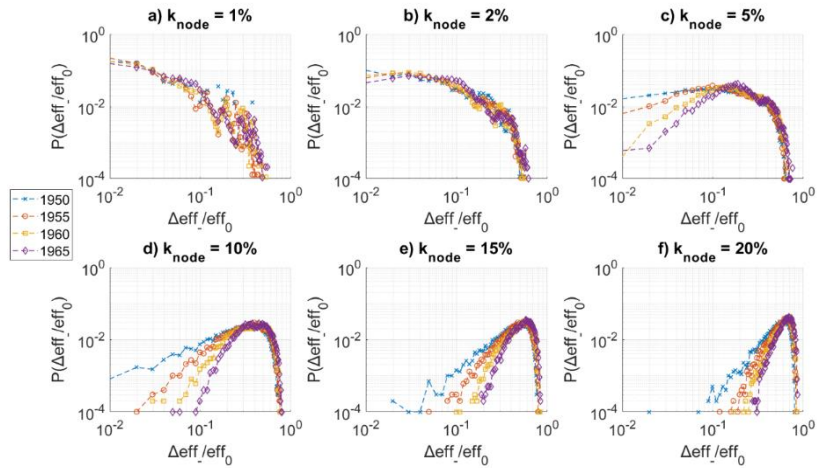

Figure 4. Decrease of network efficiency given k% of nodes removed in the case of different removals

The results for the removal of edges is shown in Table 5. While the general trends are similar to that of node removals, there are a number of differences to recognize. Given the tree-like structure of the network, the removal of a low percentage of edges causes limited damage, approx. 60-70% of the removal of the same percentage of nodes. Also, above 10% there is no significant change in the ratio of damages between 1950–1965, which again indicates that while the evolutionary process resulted in a larger network, did not enhance the vulnerability considerably. Pdfs of the results are presented in Figure 5. in detail.

Table 5. Maximal values of $\Delta eff./eff_0$ given k% of edges removed

| Year | 1% | 2% | 5% | 10% | 15% | 20% |
|------|------|------|------|------|------|------|
| 1950 | 0.2318 | 0.4655 | 0.5539 | 0.6763 | 0.7591 | 0.7892 |
| 1965 | 0.3821 | 0.4591 | 0.6398 | 0.7293 | 0.8064 | 0.8624 |
| 1965/1950 | 1.6483 | 0.9863 | 1.1551 | 1.0784 | 1.0623 | 1.0927 |



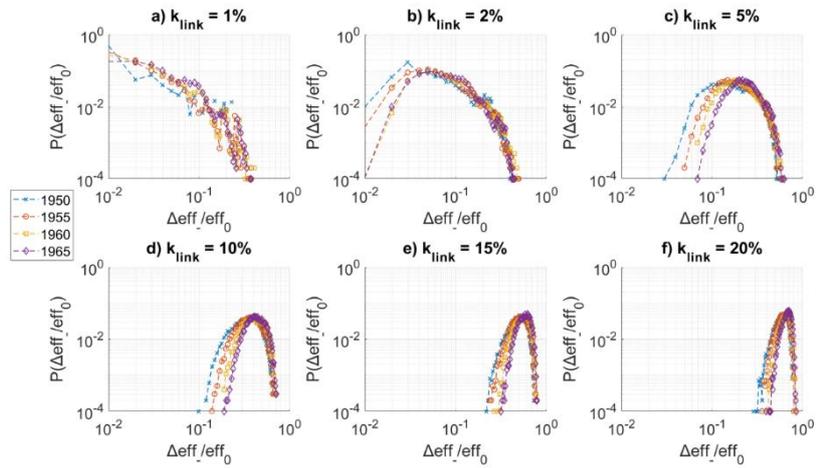

Figure 5. Decrease of network efficiency given k% of nodes removed in the case of different removals

### 3.4. Changes of centrality measures

Given that the literature shows a high correlation between the value of the damage caused by the removal of given nodes/edges and their position in the topology, we analysed the changes of centrality measures as well.

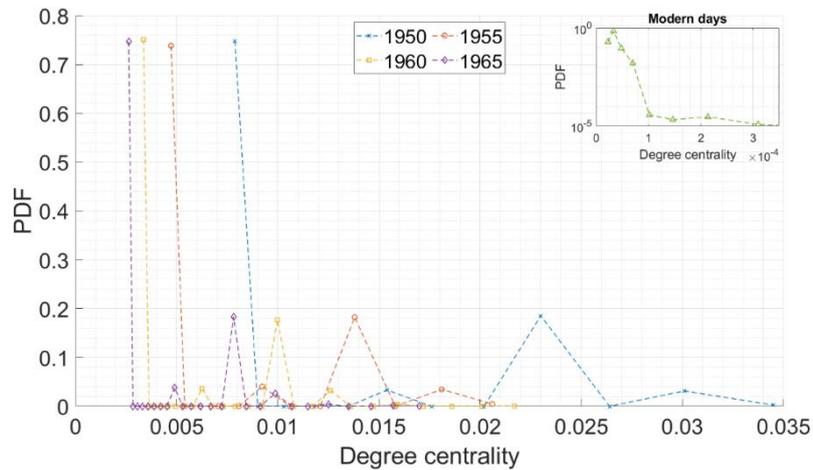

Figure 6. Probability density function of degree centrality for historic and modern network topologies



Figure 6. shows evolution of the degree centrality distributions between 1950–1965. As one may observe, the modes of distribution shift to the left (to smaller values) and the shape of the distribution becomes more unimodal. This may be due to the expansion of the networks; as more and more settlements were connected to the main lines the structure of the network has evolved becoming more tree-like in structure. This is also reflected in the pdf of the modern days network, where a steep drop is seen.

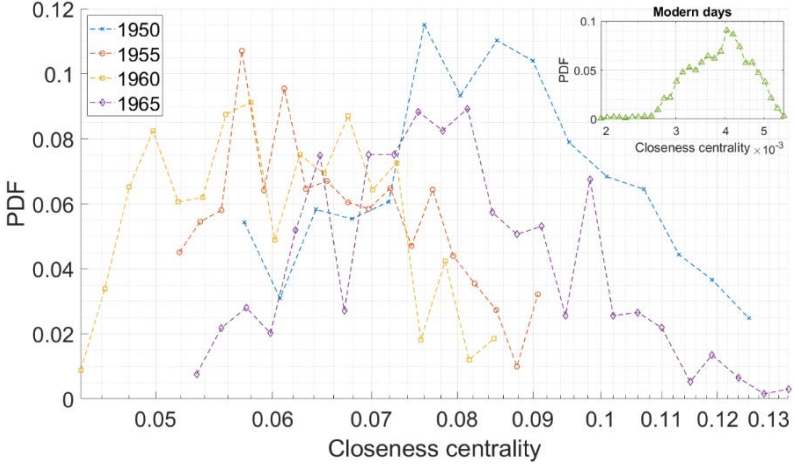

Figure 7.  Probability density function of closeness centrality for historic and modern network topologies

Closeness centrality distributions were expected to shift to smaller values simply because of the definition of CC as an inverse of the spatial extent; the average shortest distance increases with continued spatial growth. However, we observe that for the 1965 network the pdf seemed to have shifted to the right which almost resembles that of the 1950 network. The possible explanation for this result is the same as mentioned previously, the extensive 35 kV development, which connected northern and southern parts of the region and thus significantly altered the spatial importance of certain nodes.



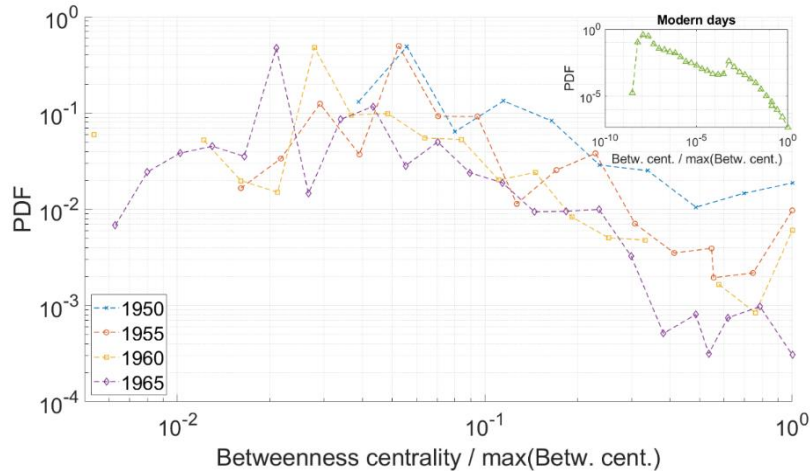

Figure 8. Probability density function of betweenness centrality for historic and modern network topologies

Because of its nature as a planar graph, the growth the DÉDÁSZ distribution network is also subjected to the conditions of planarity, as in the case of road networks [25]. The expansion of the examined network resulted to peripheral nodes found at the actual peripheries evident in the increased probability of finding nodes with the very low betweenness centralities. At the earlier state of the evolution, we can observe significant peaks at the tails of the 1950–1960 networks (Figure 8.). However, for the 1965 and modern-day network, we observe the tails to decline which signals decrease of highly-between nodes. Additionally, as the years progress, the shape of the historical probability distributions approach the state of the 'modern days' network which can be characterized as a sequence of peaking and declining. To draw an analogy, the peaking was likely caused by the fact that power lines were constructed from a limited number of central locations and after major lines were built, the importance of these locations became smaller, thus also decreasing their BC. We note that the probability distribution of the betweenness centrality for the modern day network follows the same universal shape as random planar graphs as in the case of road networks [26][27].

*3.5. Discussion*

Based on the results presented in sections 3.1–3.4, the main structural properties and thus the vulnerability of the DÉDÁSZ network shows only minor changes during the examined period, despite



its size increasing from 138 nodes and 144 edges to 400 nodes and 417 edges. Clustering remains low throughout the examined period, and this, together with the relatively high value of the average path length, describes a predominantly tree-like topology. Therefore, neither the small-world property nor the scale-free node degree distribution appear, which are usually features of developing and thus growing networks. This result confirms our described hypothesis that the increase in size is mainly not caused by an evolutionary process, but rather by the connection of geographically (and thus spatially) constrained supply areas. At the same time, since identical evolutionary processes took place within these areas, the nature of the network's vulnerability did not change as a result of 20 years of development.

To examine to what extent early developments affected the modern structure of the network, we displayed the nodes belonging to the 95th percentile for all three centrality metrics. Considering degree centrality (Figure 8.), the nodes with the largest connecting edges are typically located in large settlements, like Kaposvár, Nagykanizsa, Pécs and Szigetvár. (Local power plants of these settlements date back to 1893, 1892, 1894 and 1927, respectively.) It can also be seen that some nodes have lost their 'importance' over time; these typically belong to industrial facilities, which have been closed due to obsolescence. The pattern of nodes with the highest closeness centrality values once again underline the spatial characteristics of the supply area (Figure 9.); the only major change between 1965 and the modern days is that due to the far larger numbers, the nodes belonging to the 95th percentile are even more concentrated geographically. However, cities of Dombóvár, Kaposvár, Nagyatád, Szigetvár and Tamási are prominent in both maps. Looking at the nodes with the highest betweenness centrality (Figure 10.), we see that these are aligned along the 35 kV network of the region. If we also compare this with the 1945 status of the network (Figure 1.) it is clearly visible that high betweenness value nodes practically move on a 'trajectory', following the extension of the 35 kV lines. (In 1950 on the Máza–Pécs route, in 1955 on the Máza–Pécs–Szigetvár–Nagyatád route, in 1960 on the Máza–Pécs–Szigetvár–Nagyatád–Nagykanizsa route, and finally in 1965 on the Szigetvár–Nagyatád–Nagykanizsa–Kaposvár formation.) Accepting the consensus of the literature



that nodes with the highest betweenness centrality value are the most vulnerable parts of the network, network developments of the 1950–1965 period did not significantly decrease vulnerability, as they mostly focused on connecting the settlements to the main lines.

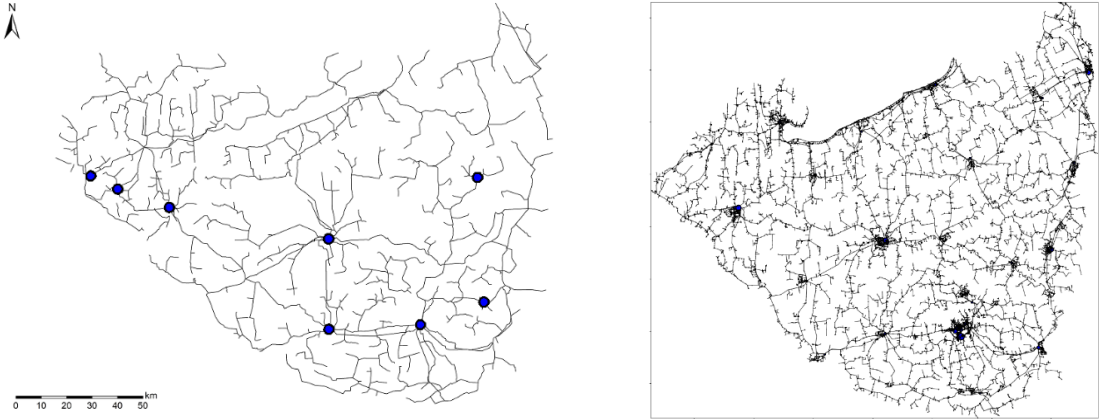

Figure 8. Nodes belonging to the 95th percentile, degree centrality in 1965 (left) and modern days (right)

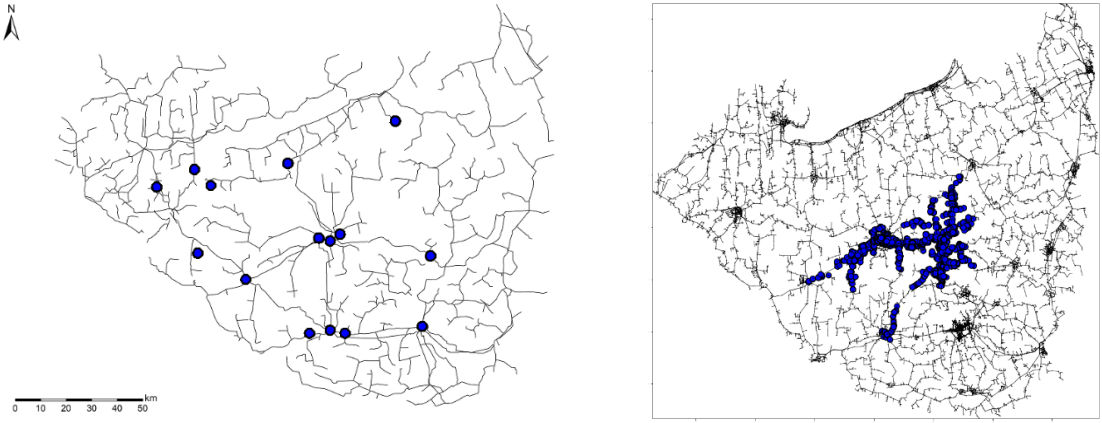

Figure 9. Nodes belonging to the 95th percentile, closeness centrality in 1965 (left) and modern days (right)



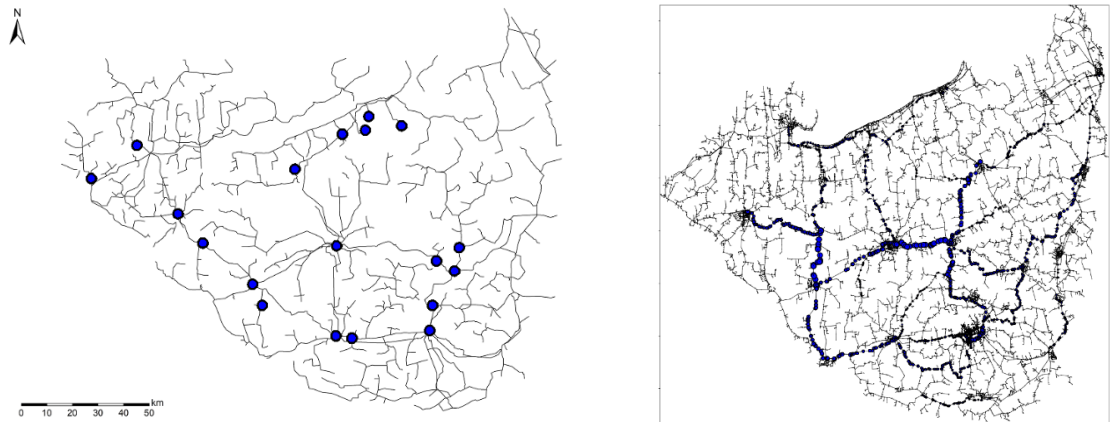

Figure 10. Nodes belonging to the 95th percentile, betweenness centrality in 1965 (left) and modern days (right)

### 4. Conclusions

In this paper the vulnerability of a regional distribution network was studied during its evolutionary period, using a complex network approach. Topological efficiency was calculated against node and edge removals of different proportions, between 1 and 20%. The results show that the tolerance of the evolving grid against such disturbances remained largely unchanged during the examined period. This implies that the increase in size is dominantly caused by the connection of geographically and spatially constrained supply areas and not by an evolutionary process, as one would expect knowing the origin of the data. We also found that the structural characteristics of the network were formed by two different processes, the electrification of settlements and the interconnection of industrial areas. Finally, we compared the spatial distribution of various centrality metrics to see to what extent historic topologies are resembled in modern-days'.

While taking into consideration the simplifications, which were necessary to create the historic network models, we can say that the vulnerability of the examined distribution network was determined by the technical and economic decisions related to industrial electrification, and this leaves its mark even 60–70 years later in an inherent way. The importance of this observation is given by the fact that the development of electrical networks usually means expansion, and the termination of power lines or transformer stations only occurs as an exception. As a result, the



construction of new infrastructure will have an impact spanning decades, most likely also for periods when the power system will operate according to different principles.

However, this apparent disadvantage can also be turned into an advantage: by understanding the development models and their impact on vulnerability, we can significantly support the redesign necessary due to the ongoing green transition, the emergence of reconfigurable topologies and, above all, the supply of consumers.

**CRediT authorship contribution statement**

**Bálint Hartmann:** Writing – original draft, Visualization, Validation, Supervision, Resources, Project administration, Methodology, Investigation, Funding acquisition, Formal analysis, Conceptualization. **Tamás Soha:** Writing – original draft, Visualization, Validation. **Michelle T. Cirunay:** Writing – original draft, Formal analysis, Data curation. **Tímea Erdei:** Writing – original draft, Data curation.


**Acknowledgement**

Bálint Hartmann acknowledges the support of the Bolyai János Research Scholarship of the Hungarian Academy of Sciences and the support of the ÚNKP-23-5-BME-438 New National Excellence Program of the Ministry for Culture and Innovation from the source of the National Research, Development and Innovation Fund.